# Bounded Independence Fools Halfspaces

Ilias Diakonikolas[*]        Parikshit Gopalan[†]        Ragesh Jaiswal[‡]
Columbia University        MSR-Silicon Valley        Columbia University

Rocco A. Servedio[§]        Emanuele Viola[¶]
Columbia University        Northeastern University

October 22, 2018

## Abstract

We show that any distribution on $\{-1, +1\}^n$ that is $k$-wise independent fools any halfspace $h : \{-1, +1\}^n \to \{-1, +1\}$ with error $\epsilon$ for $k = O(\log^2(1/\epsilon)/\epsilon^2)$. Up to logarithmic factors, our result matches a lower bound by Benjamini, Gurel-Gurevich, and Peled (2007) showing that $k = \Omega(1/(\epsilon^2 \cdot \log(1/\epsilon)))$. Using standard constructions of $k$-wise independent distributions, we obtain the first explicit pseudorandom generators $G : \{-1, +1\}^s \to \{-1, +1\}^n$ that fool halfspaces. Specifically, we fool halfspaces with error $\epsilon$ and seed length $s = k \cdot \log n = O(\log n \cdot \log^2(1/\epsilon)/\epsilon^2)$.

Our approach combines classical tools from real approximation theory with structural results on halfspaces by Servedio (Comput. Complexity 2007).

---

[*]Research supported by NSF grants CCF-0728736, CCF-0525260, and by an Alexander S. Onassis Foundation Fellowship. Email: `ilias@cs.columbia.edu`

[†]Email: `parik@microsoft.com`

[‡]Research supported by DARPA award HR0011-08-1-0069. Email: `rjaiswal@cs.columbia.edu`

[§]Supported in part by NSF grants CCF-0347282, CCF-0523664 and CNS-0716245, and by DARPA award HR0011-08-1-0069. Email: `rocco@cs.columbia.edu`

[¶]This work was partially done while the author was a postdoctoral fellow at Columbia University, supported by grants NSF award CCF-0347282 and NSF award CCF-0523664. Email: `viola@ccs.neu.edu`

# 1 Introduction

*Halfspaces*, or threshold functions, are a central class of Boolean functions $h : \{-1, +1\}^n \to \{-1, +1\}$ of the form:

$$h(x) = \text{sign}(w_1 x_1 + \cdots + w_n x_n - \theta),$$

where the weights $w_1, \ldots, w_n$ and the threshold $\theta$ are arbitrary real numbers. These functions have been studied extensively in a variety of contexts. In computer science, the work on halfspaces dates back to the study of switching functions, see for instance the books [Der65, Hu65, LC67, She69, Mur71]. In computational complexity, much effort has been put into understanding constant-depth circuits of halfspaces. On the one hand this has resulted in surprising inclusions (such as the simulation of depth-$d$ circuits of halfspaces by depth-$(d + 1)$ circuits of majority gates [GHR92, GK98]), but on the other hand many seemingly basic questions remain unsolved: for instance it is conceivable that every function in NP is computable by a depth-2 circuit of halfspaces [HMP+93, Kra91, KW91, FKL+01]. In learning theory, the problem of learning an unknown halfspace has arguably been the most influential problem in the development of the field, with algorithms such as Perceptron, Weighted Majority, Boosting, and Support Vector Machines emerging from this study. Halfspaces (with non-negative weights) have also been studied extensively in game theory and social choice theory, where they are referred to as "weighted majority games" and have been analyzed as models for voting, see e.g., [Pen46, Isb69, DS79, TZ92].

In this work we make progress on a natural complexity-theoretic question about halfspaces. We construct the first explicit pseudorandom generators $G : \{-1, +1\}^s \to \{-1, +1\}^n$ with short seed length $s$ that fool any halfspace $h : \{-1, +1\}^n \to \{-1, +1\}$, i.e. satisfy

$$|\mathbf{E}_{x \in \{-1,+1\}^s}[h(G(x))] - \mathbf{E}_{x \in \{-1,+1\}^n}[h(x)]| \leq \epsilon,$$

for a small $\epsilon$. We actually prove that the class of distributions known as $k$-wise independent has this "fooling" property for a suitable $k$; as pointed out below, a generator can then be obtained using any of the standard explicit constructions of such distributions.

**Definition 1.1.** A distribution $\mathcal{D}$ on $\{-1, +1\}^n$ is $k$-*wise independent* if the projection of $\mathcal{D}$ on any $k$ indices is uniformly distributed over $\{-1, +1\}^k$.

**Theorem 1.2** (Main). Let $\mathcal{D}$ be a $k$-wise independent distribution on $\{-1, +1\}^n$, and let $h : \{-1, +1\}^n \to \{-1, +1\}$ be a halfspace. Then $\mathcal{D}$ fools $h$ with error $\epsilon$, i.e.,

$$|\mathbf{E}_{x \leftarrow \mathcal{D}}[h(x)] - \mathbf{E}_{x \leftarrow \mathcal{U}}[h(x)]| \leq \epsilon, \qquad \text{provided} \qquad k \geq \frac{C}{\epsilon^2} \log^2\left(\frac{1}{\epsilon}\right),$$

where $C$ is an absolute constant and $\mathcal{U}$ is the uniform distribution over $\{-1, +1\}^n$.

Our Theorem 1.2 matches up to logarithmic factors a lower bound by Benjamini, Gurel-Gurevich and Peled [BGGP07] establishing that if $k = o(1/(\epsilon^2 \cdot \log(1/\epsilon)))$ then there exists a $k$-wise independent distribution $\mathcal{D}$ on $\{-1, +1\}^n$ such that $\text{Pr}_{x \leftarrow \mathcal{D}}[\sum_i x_i \geq 0] > 1/2 + \epsilon$, i.e. $\mathcal{D}$ does not fool the majority function with error $\epsilon$.

Standard explicit constructions of $k$-wise independent distributions over $\{-1, +1\}^n$ have seed length $O(k \cdot \log n)$ [CG89, ABI86], which is optimal up to constant factors [CGH+85]. Plugging these in Theorem 1.2, we obtain explicit pseudorandom generators $G : \{-1, +1\}^s \to \{-1, +1\}^n$ that fool any halfspace $h : \{-1, +1\}^n \to \{-1, +1\}$ with error $\epsilon$ and have seed length $s = O(\log n \cdot \log^2(1/\epsilon)/\epsilon^2)$.

**Background and comparison with previous explicit generators.** The literature is rich with explicit generators for various classes, such as small constant-depth circuits with various gates [AW, Nis91, LVW93, Vio07, Baz07, Bra09], low-degree polynomials [NN93, AGHP92, BV07, Lov08, Vio08], and one-way small-space algorithms [Nis92]. Many of these classes (such as low-degree polynomials and $AC^0$ circuits) provably cannot implement halfspaces, and it is not known how to implement an arbitrary halfspace in any of these classes, so none of these results gives Theorem 1.2. However, some of these results [Nis92, LVW93, Vio07] give generators for the *restricted class* of halfspaces given by $h(x) = \text{sign}(\sum_{i=1}^n w_i x_i - \theta)$ where the weights are integers of magnitude at most poly($n$). While it is well known that every halfspace has a representation with integer weights, in general it is not possible to represent an arbitrary halfspace with poly($n$) integer weights. Indeed, an easy counting argument (see e.g. [MT94, Hås94]) shows that if the weights are required to be integers then almost all halfspaces require weights of magnitude $2^{\Omega(n)}$, and in fact some halfspaces require weights of magnitude $2^{\Theta(n \log n)}$ [Hås94]. Our result is for the entire class of halfspaces with no restriction on the weights, and much of the richness of halfspaces only comes in this setting; for example, the "odd-max-bit" function [Bei94], the "universal halfspace" [GHR92], and other important halfspaces [Hås94] all require exponentially large integer weights. Moreover, even for the restricted class of halfspaces where the weights are integers of magnitude at most poly($n$), previous techniques [Nis92] give seed length $s = O(\log^2 n)$ at best, while we achieve $s = O(\log n)$ for constant error.

**Other related results.** Several recent papers have studied the power of $k$-wise independent distributions. An exciting recent result of Braverman [Bra09], which builds on an earlier breakthrough of Bazzi [Baz07] (simplified by Razborov [Raz08]), shows that polylog($n$)-wise independent distributions fool small constant-depth circuits, settling a conjecture of Linial and Nisan [LN90]. Benjamini *et al.* [BGGP07] showed that any $O(1/\epsilon^2)$-wise independent distribution $\mathcal{D}$ on $\{-1, +1\}^n$ satisfies $|\Pr_{x \leftarrow \mathcal{D}}[\sum_i x_i \geq 0] - 1/2| \leq \epsilon$, i.e., such distributions fool the majority function. (We discuss [BGGP07] in more detail shortly. Here we note that their result does not seem immediately relevant for constructing generators, because to fool the majority function, with optimal error 0, one can just output $1^n$ with probability $1/2$ and $(-1)^n$ with probability $1/2$.) None of these results applies to general halfspaces.

The problem of constructing pseudorandom generators for halfspaces has been considered by several authors in the recent literature. Rabani and Shpilka give an explicit construction of an $\epsilon$-net, or $\epsilon$-hitting set, for halfspaces [RS08]: a set of size poly($n, 1/\epsilon$) which is guaranteed to contain at least one point where $h(x) = +1$ and at least one point where $h(x) = -1$ for any halfspace $h$ which takes on both values with probability at least $\epsilon$ under the uniform distribution. However, their construction does not offer any guarantees about the distribution of these values. [RS08] pose as a research goal "to build methodically a theory of pseudorandom generators for geometric functions" such as halfspaces.

The problem of pseudorandom generators for halfspaces also arose in recent work by Gopalan and Radhakrishnan [GR09] on finding duplicates in a data stream. They required a pseudorandom distribution that allows one to estimate the influence of a variable in a halfspace, a problem which is in fact equivalent to constructing a pseudorandom generator for a related halfspace. They observe that Nisan's space generator [Nis92] suffices for the halfspaces arising in their context, but they raise the problem of constructing pseudorandom generators for general halfspaces. Our result does not improve the space bounds for their problem, but it does make the analysis simpler.

## 1.1   Techniques

Our proof combines tools from real approximation theory with structural results regarding halfspaces. An important notion is that of an *$\epsilon$-regular* halfspace; which is a halfspace $h(x) = \text{sign}(\sum_i w_i x_i - \theta)$ where no more than an $\epsilon$-fraction of the 2-norm of its coefficient vector $(w_1, \ldots, w_n)$ comes from any single coefficient $w_i$. We first show that $k$-wise independence fools all $\epsilon$-regular halfspaces, and then use this to prove that $k$-wise independence fools all halfspaces. Our proof can be broken down conceptually into three steps.

**Step 1: Fooling regular halfspaces.**   Our starting point is Bazzi's observation ([Baz07], Theorem 4.2) that to establish that every $k$-wise independent distribution on $\{-1, +1\}^n$ fools a Boolean function $f : \{-1, +1\}^n \to \{-1, +1\}$ with error $\epsilon$, it is sufficient to exhibit two "sandwiching" polynomials $q_\ell, q_u : \{-1, +1\}^n \to \{-1, +1\}$ of degree at most $k$ such that:

- $q_u(x) \geq f(x) \geq q_\ell(x)$ for all $x \in \{-1, +1\}^n$; and

- $\mathbf{E}_{\mathcal{U}}[q_u(x) - f(x)], \mathbf{E}_{\mathcal{U}}[f(x) - q_\ell(x)] \leq \epsilon$.

Using classical tools from real approximation theory, we give a self-contained proof of the existence of univariate polynomials of degree $K(\epsilon) := \tilde{O}(1/\epsilon^2)$ which, roughly speaking, provide a good sandwich approximator to the *univariate* function $\text{sign}(t)$ *under the normal distribution on* $\mathbf{R}$. This is useful for us because of the following simple but crucial insight: for any regular halfspace $h(x) = \text{sign}(w \cdot x - \theta)$, the argument $w \cdot x - \theta$ is well-approximated by a normal random variable (a precise error-estimate for this approximation is given by the Berry-Esséen theorem). For any $\epsilon$-regular halfspace, we can thus plug $w \cdot x - \theta$ into our univariate polynomials, and obtain low-degree sandwich polynomials for $h$. This establishes that $K(\epsilon)$-wise independence fools all $\epsilon$-regular halfspaces.

Of course, there are halfspaces $\text{sign}(w \cdot x - \theta)$ that are far from being $\epsilon$-regular and have $w \cdot x - \theta$ distributed very unlike a Gaussian. To tackle general halfspaces, we use the notion of the *$\epsilon$-critical index of a halfspace*, which was (implicitly) introduced in [Ser07] and has since played a useful role in several recent results on halfspaces [OS08, MORS09, DS09]. Briefly, assuming that the weights $w_1, \ldots, w_n$ are sorted by absolute value, the $\epsilon$-critical index is the first index $\ell$ so that the weight vector $(w_\ell, w_{\ell+1}, \ldots, w_n)$ is $\epsilon$-regular. The previous Step 1 handled halfspaces that are regular, corresponding to $\ell = 1$. We now proceed by analyzing two cases, based on whether or not $1 < \ell < L(\epsilon)$, or $\ell \geq L(\epsilon)$, for $L(\epsilon) := \tilde{O}(1/\epsilon^2)$. In both cases, it is convenient to think of the variables as partitioned into a "head" part consisting the first $L(\epsilon)$ variables and corresponding to the largest weights, and of a "tail" part consisting of the rest.

**Step 2: Fooling halfspaces with small critical Index ($\ell < L(\epsilon)$).** We argue that for every setting of the head variables, the $\epsilon$-regularity of the tail is sufficient to ensure that the overall halfspace gives the right bias. More precisely, we assume that our distribution $\mathcal{D}$ is $(K(\epsilon) + L(\epsilon))$-wise independent, and note that each setting of the $\ell$ head variables gives an $\epsilon$-regular halfspace $\text{sign}(w \cdot x - \theta')$ over the tail variables (with the constant $\theta'$ depending on the values of the head variables). Since the marginal distribution on the tail variables is $K(\epsilon)$-wise independent for every setting of the head variables, the distribution $\mathcal{D}$ fools all such halfspaces.

**Step 3: Fooling halfspaces with large critical index ($\ell \geq L(\epsilon)$).** In this case, we show that the setting of the head variables alone is very likely to determine the value of the function. More precisely, we show that a uniform random assignment to the head variables is very likely to yield a halfspace $\text{sign}(w_T \cdot x_T - \theta')$ over the tail variables $T$ in which

$$|\theta'| > \|w_T\|_2/\epsilon. \qquad (\star)$$

Now, as long as the tail variables are pairwise independent, by Chebyshev's inequality it follows that the value $w_T \cdot x_T$ will be sharply concentrated within $[-\|w_T\|_2, +\|w_T\|_2]$. So, for most settings of the head variables, we get something very close to a constant function over the tail variables. Since a $(K(\epsilon) + 2)$-wise independent distribution gives us uniform randomness for the head variables and pairwise independence for the tail variables, bounded independence fools these halfspaces as well.

The key idea behind the proof of $(\star)$ is that up to the critical index $\ell$ – which in this case is large ($\ell \geq L(\epsilon)$) – the weights $(w_1, \ldots, w_{\ell-1})$ must be decreasing fairly rapidly; this allows us to prove strong anti-concentration for the distribution of $\theta'$, which in turn yields $(\star)$.

Overall, the amount of independence required for all the three steps to work is:

$$\max\{K(\epsilon), K(\epsilon) + L(\epsilon), K(\epsilon) + 2\} = \tilde{O}(1/\epsilon^2),$$

concluding this sketch of the proof of Theorem 1.2.

**Univariate approximations to the sign function.** As mentioned above, our approach relies on the existence of low-degree univariate sandwich approximators to the sign function under the normal distribution on $\mathbf{R}$. Low-degree approximations to the sign function have been studied in both computer science and mathematics (see for instance [Pat92, EY07, KS07] and the references therein). However it appears that these results do not fit all our requirements. Below we discuss how our approach relates to the work of Benjamini *et al.* [BGGP07] and Eremenko and Yuditskii [EY07].

Benjamini *et al.* prove that $O(1/\epsilon^2)$-wise independence suffices to fool the majority function, using machinery from the theory of the classical moment problem. However, their proof seems to be tailored quite specifically to the majority function, where the moments can be understood in terms of Krawtchouk polynomials and known bounds on such polynomials can be applied, so it seems difficult to extend their approach to general halfspaces (or indeed even to slight variants of the majority function).

Bazzi's condition on the existence of degree-$k$ sandwiching polynomials mentioned above is in fact both necessary and sufficient for all $k$-wise independent distributions to fool a function $f$. Thus the [BGGP07] theorem implies the existence of $O(1/\epsilon^2)$-degree multivariate sandwich polynomials for the majority function; symmetrization then implies that there exist univariate polynomials which, roughly speaking, provide good sandwich approximation to the function $\text{sign}(t)$ under the binomial distribution. This is similar in spirit to the result we establish (mentioned in Step 1 above) about univariate polynomial approximators, but there is a crucial difference: since the binomial distribution is supported only on the integers $\{-n, \ldots, n\}$, it seems difficult to infer much about the behavior of the univariate polynomial implicit in [BGGP07] on values outside of $\{-n, \ldots, n\}$. Hence, it is unclear whether these polynomials can be used for general (or even regular) halfspaces.

In contrast, we work with the *best possible* pointwise approximation to the function $\text{sign}(t)$ on the (piecewise) *continuous* domain $[-1, -a] \cup [a, 1]$. This uniform error bound is convenient for dealing with regular halfspaces; moreover, working with the optimal pointwise approximator allows us to exploit various properties of optimal approximators that follow from the theory of Chebyshev approximation, in a way that is crucial for us to obtain the required "univariate sandwich approximators."

We note that a recent work in approximation theory [EY07] analyzes the error achieved by this optimal polynomial and in particular establishes the limiting behavior of the error. For our purposes, though, we require the error to converge to the limit fairly rapidly and it is unclear whether the results of [EY07] guarantee this. We present an error analysis which is elementary (it only uses basic approximation theory) and moreover matches the limiting bounds of [EY07] up to a constant factor.

**Organization.** In Section 3 we show how a certain univariate polynomial approximator to $\text{sign}(t)$ yields low-degree sandwich polynomials for $\epsilon$-regular halfspaces over $\{-1, 1\}^n$. In Section 4 we construct the required univariate polynomial, which essentially gives sandwich polynomials for $\text{sign}(t)$ under the normal distribution. In Section 5 we show how non-regular halfspaces can be fooled using our results for regular halfspaces, concluding the proof or our main theorem.

# 2    Preliminaries

Recall that the univariate function sign$(t)$ takes value $+1$ for $t \geq 0$ and $-1$ for $t < 0$.

**Definition 2.1** (Halfspace). A *halfspace* is a Boolean function $f : \{-1, +1\}^n \to \{-1, +1\}$ which can be expressed as $f(x) = \text{sign}(\sum_i w_i x_i - \theta)$ for some $\theta \in \mathbf{R}$, $(w_1, \ldots, w_n) \in \mathbf{R}^n$.

Throughout this paper we assume without loss of generality that halfspaces are normalized to satisfy $w_1^2 + \cdots + w_n^2 = 1$. Such a representation can always be obtained by appropriate scaling.

**Definition 2.2** (Fooling a Function Class). Let $f : \{-1, +1\}^n \to \{-1, +1\}$ be any function. We say that a distribution $\mathcal{D}$ over $\{-1, +1\}^n$ *fools* $f$ *with error* $\epsilon$, or $\epsilon$-*fools* $f$, if

$$|\mathbf{E}_{x \leftarrow \mathcal{D}}[f(x)] - \mathbf{E}_{x \leftarrow \mathcal{U}}[f(x)]| \leq \epsilon,$$

where $\mathcal{U}$ denotes the uniform distribution over $\{-1, +1\}^n$. We say that $\mathcal{D}$ fools a class of functions $\mathcal{F}$ if $\mathcal{D}$ fools every $f \in \mathcal{F}$.

We require a few basic facts from probability theory: the Berry-Esséen theorem and the standard tail bounds of Hoeffding and Chebyshev. We discuss them next.

The Berry-Esséen theorem is a version of the Central Limit Theorem with explicit error bounds:

**Theorem 2.3.** *(Berry-Esséen) Let $X_1, \ldots, X_n$ be a sequence of independent random variables satisfying $\mathbf{E}[X_i] = 0$ for all $i$, $\sqrt{\sum_i \mathbf{E}[X_i^2]} = \sigma$, and $\sum_i \mathbf{E}[|X_i|^3] = \rho_3$. Let $S = (X_1 + \cdots + X_n)/\sigma$ and let $F$ denote the cumulative distribution function (cdf) of $S$. Then*

$$\sup_x |F(x) - \Phi(x)| \leq C\rho_3/\sigma^3,$$

*where $\Phi$ is the cdf of a standard Gaussian random variable (with mean zero and variance one), and $C$ is a universal constant. [Shi86] has shown that one can take $C = .7915$.*

**Corollary 2.4.** *Let $x_1, \ldots, x_n$ denote independent uniformly $\pm 1$ random signs and let $w_1, \ldots, w_n \in \mathbf{R}$. Write $\sigma = \sqrt{\sum_i w_i^2}$, and assume $|w_i|/\sigma \leq \tau$ for all $i$. Then for any interval $[a, b] \subseteq \mathbf{R}$,*

$$\left| \Pr[a \leq w_1 x_1 + \cdots + w_n x_n \leq b] - \Phi([\tfrac{a}{\sigma}, \tfrac{b}{\sigma}]) \right| \leq 2\tau,$$

*where $\Phi([c, d]) := \Phi(d) - \Phi(c)$. In particular,*

$$\Pr[a \leq w_1 x_1 + \cdots + w_n x_n \leq b] \leq \frac{|b - a|}{\sigma} + 2\tau.$$

For completeness we recall the Hoeffding and Chebyshev bounds:

**Theorem 2.5** (Hoeffding). *Fix any $w \in \mathbf{R}^n$. For any $\gamma > 0$, we have*

$$\Pr_{x \leftarrow \mathcal{U}}[w \cdot x \geq \gamma \|w\|] \leq e^{-\gamma^2/2} \quad \text{and} \quad \Pr_{x \leftarrow \mathcal{U}}[w \cdot x \leq -\gamma \|w\|] \leq e^{-\gamma^2/2}.$$

**Theorem 2.6** (Chebyshev). *For any random variable $X$ with $\mathbf{E}[X] = \mu$ and $\text{Var}[X] = \sigma^2$ and any $k > 0$,*

$$\Pr[|X - \mu| \geq k\sigma] \leq \frac{1}{k^2}.$$

# 3   Fooling regular halfspaces

In this section we show how to fool regular halfspaces, defined next (recall all our halfspaces are normalized to satisfy $w_1^2 + \cdots + w_n^2 = 1$).

**Definition 3.1** (Regular Halfspace). A halfspace $f$ is said to be $\epsilon$-*regular* if it can be expressed as $f(x) = \text{sign}(w \cdot x - \theta)$ where for all $i = 1, \ldots, n$, we have

$$|w_i| \leq \epsilon.$$

An $\epsilon$-regular halfspace $f(x) = \text{sign}(w \cdot x - \theta)$ has the convenient property that the cumulative distribution function (cdf) of $w \cdot x - \theta$ is everywhere within $\pm O(\epsilon)$ of the cdf of the shifted Gaussian $N(-\theta, 1)$. This is a direct consequence of the Berry-Esséen Theorem.

Given $\epsilon > 0$, we define the following parameters:

$$a(\epsilon) := \frac{\epsilon^2}{C \log(1/\epsilon)},$$

$$K(\epsilon) := \frac{4c \log(1/\epsilon)}{a} + 2 < \frac{5c}{a} \log(1/\epsilon) = O\left(\log^2(1/\epsilon)/\epsilon^2\right).$$

We assume without loss of generality that $\epsilon$ is a sufficiently small power of 2 (i.e., $\epsilon = 2^{-i}$ for some integer $i$). The positive constants $C$ and $c$ will be chosen later; but (with foresight), we will require that $C \gg c$.

In this section we prove the following:

**Theorem 3.2** (Fooling $\epsilon$-regular halfspaces). *Any $K(\epsilon)$-wise independent distribution fools $\epsilon$-regular halfspaces with error $12\epsilon$.*

To prove the theorem we construct certain "sandwiching" polynomials. We now define such polynomials and then explain why they are sufficient for our purposes.

**Definition 3.3.** Let $f : \{-1, +1\}^n \to \{-1, +1\}$ be a Boolean function. A pair of real-valued polynomials $q_\ell(x_1, \ldots, x_n)$, $q_u(x_1, \ldots, x_n)$ are said to be $\epsilon$-*sandwich polynomials of degree $k$ for $f$* if they have the following properties:

- $\deg(q_u), \deg(q_\ell) \leq k$;

- $q_u(x) \geq f(x) \geq q_\ell(x)$ for all $x \in \{-1, +1\}^n$;

- $\mathbf{E}_{x \leftarrow \mathcal{U}}[q_u(x) - f(x)] \leq \epsilon$ and $\mathbf{E}_{x \leftarrow \mathcal{U}}[f(x) - q_\ell(x)] \leq \epsilon$.

The following fact relates sandwiching polynomials to fooling:

**Lemma 3.4** (Bazzi). *Let $f : \{-1, +1\}^n \to \{-1, +1\}$ be a Boolean function. Every $k$-wise independent distribution $\epsilon$-fools $f$ if and only if there exist $\epsilon$-sandwich polynomials of degree $k$ for $f$.*

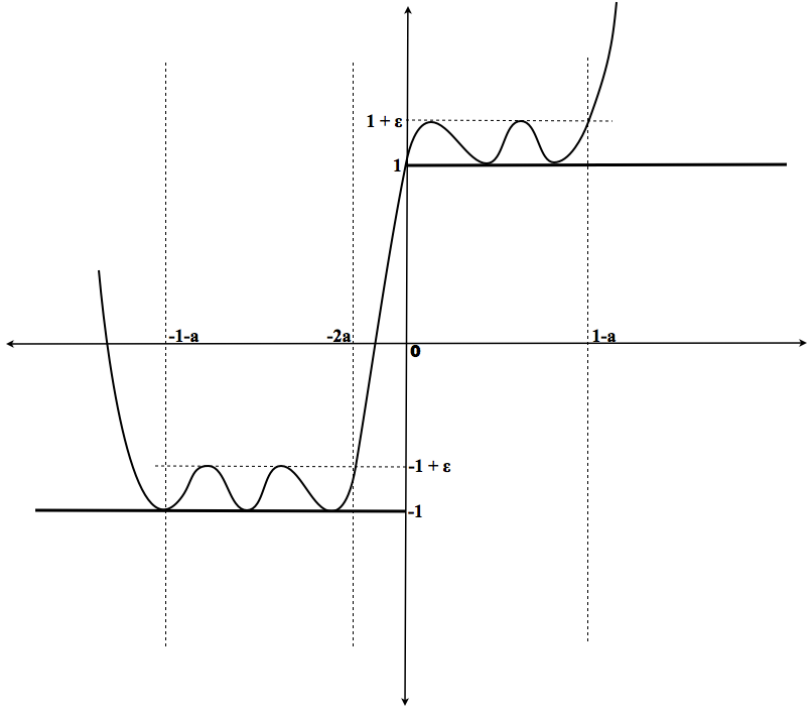

Figure 1: Qualitative plot of polynomial $P$.

We only use the "if" direction of this lemma for our proof, which follows straightforwardly by linearity of expectation. The other direction is a consequence of LP–duality (see [Baz07] for a proof).

To construct appropriate sandwiching polynomials, we start by exhibiting a *univariate* polynomial $P : \mathcal{R} \to \mathcal{R}$ that approximates the function sign $: \mathcal{R} \to \{-1, 1\}$ in a certain specialized way which we discuss shortly. Let us be given an $\epsilon$-regular halfspace $h(x) = \text{sign}(w \cdot x - \theta)$, and assume that $|\theta|$ is small (the case where $|\theta|$ is large is simpler). We obtain the upper sandwiching polynomial $q_u$ in Definition 3.3 by plugging into $P$ the value $w \cdot x - \theta$ scaled by a large $Z \geq 0$:

$$q_u(x) := P\left(\frac{w \cdot x - \theta}{Z}\right).$$

For the lower sandwiching polynomial $q_\ell$ we will use $-P(-t)$. The key properties of $P(t)$ are that (1) $P(t) \geq \text{sign}(t)$ for every $t \in \mathcal{R}$, (2) $P(t)$ gives a good (error $\epsilon$) pointwise approximation to $\text{sign}(t)$ for $t \in [-1/2, 1/2]$ except for $t$ in the small interval $[-2a, 0]$ where the error is bounded by a constant, and (3) $P(t)$ does not grow too quickly for $|t| \geq 1/2$. For a qualitative depiction of $P$ we refer the reader to Figure 1 (this figure is not an actual plot but rather is intended to qualitatively illustrate the guarantees on the behavior of $P$ on various intervals; also the parameter $1/2$ is replaced by $1 - a \geq 1/2$ for later needs). Property (1), together with the fact that scaling by $Z$ does not change the value of the halfspace, immediately gives the sandwiching property $q_u(x) \geq h(x)$ in Definition 3.3. To establish the small-error property $\mathbf{E}_x[q_u(x) - h(x)]$ in Definition 3.3, we reason by case analysis. First, by our choice of parameters, the small interval $[-2a, 0]$ remains small even after scaling by $Z$, and so we use an anti-concentration argument to show that the input $t = (w \cdot x - \theta)/Z$ to the polynomial is unlikely to land there, and thus the contribution towards the error $\mathbf{E}_x[q_u(x) - h(x)]$ is negligible in this case. Also, whenever the input $t$ to the polynomial lands in $[-1/2, 1/2] \setminus [-2a, 0]$, the contribution to the error $\mathbf{E}_x[q_u(x) - h(x)]$ is small because by Property (2) the polynomial approximates the sign function well there: $q_u(x) - h(x) \leq \epsilon$. Finally, the event that the input $t = (w \cdot x - \theta)/Z$ to $P$ has absolute value bigger than $1/2$ corresponds to the event that $|w \cdot x - \theta| \geq Z/2$. The scaling factor $Z$ is large and the halfspace is $\epsilon$-regular, and so we can apply standard tail estimates to bound from above this probability. Since by Property (3) the polynomial $P(t)$ does not grow too quickly for $|t| \geq 1/2$, the contribution to the error $\mathbf{E}_x[q_u(x) - h(x)]$ is small even in this case.

We now proceed with the formal proof. We start with recording in the following theorem the properties of $P$.

**Theorem 3.5.** Let $0 < \epsilon < 0.1$ and let $a$ and $K$ be as defined above. There is a univariate polynomial $P(t)$ such that $\deg(P) \leq K$ with the following properties:

(1) $P(t) \geq \mathrm{sign}(t) \geq -P(-t)$ for all $t \in \mathbf{R}$;

(2) $P(t) \in [\mathrm{sign}(t), \mathrm{sign}(t) + \epsilon]$ for $t \in [-1/2, -2a] \bigcup [0, 1/2]$;

(3) $P(t) \in [-1, 1 + \epsilon]$ for $t \in (-2a, 0)$;

(4) $|P(t)| \leq 2 \cdot (4t)^K$ for all $|t| \geq 1/2$.

We defer the proof of Theorem 3.5 to Section 4 and we proceed with the proof of Theorem 3.2.

## 3.1  Proof of Theorem 3.2

Let $h(x) = \mathrm{sign}(w \cdot x - \theta)$ be an $\epsilon$-regular halfspace (and recall $w_1^2 + \cdots + w_n^2 = 1$.) Let

$$Z := \frac{\epsilon}{2a} = \frac{C \log(1/\epsilon)}{2\epsilon}.$$

We break the analysis into the following two cases, based on the magnitude of the threshold $\theta$.

### 3.1.1   $|\theta|$ is small ($|\theta| \leq Z/4$)

The sandwich polynomials we use are:

$$q_u(x) := P\left(\frac{w \cdot x - \theta}{Z}\right), \quad q_l(x) := -P\left(\frac{\theta - w \cdot x}{Z}\right). \tag{1}$$

First, observe that for every $x \in \{-1, +1\}^n$ we have

$$q_u(x) \geq h(x) \geq q_l(x).$$

This is because from Theorem 3.5 with $t = (w \cdot x - \theta)/Z$ we get

$$q_u(x) \geq \text{sign}\left(\frac{w \cdot x - \theta}{Z}\right) = \text{sign}(w \cdot x - \theta) = h(x) \geq q_l(x).$$

In the rest of this section we bound the error of the approximation.

**Lemma 3.6.** $\mathbf{E}_x[q_u(x) - h(x)] < 10\epsilon$.

*Proof.* Define the random variable $H(x) = (w \cdot x - \theta)/Z$. We prove the desired upper bound by partitioning the space into three events and bounding the contribution from each:

1. $S_1$ is the event that $H(x) \in [-\epsilon/Z, 0]$.

2. $S_2$ is the event that $|H(x)| \leq 1/2$, but $S_1$ does not happen.

3. $S_3$ is the event that $|H(x)| > 1/2$.

We have

$$\mathbf{E}_x[q_u(x) - h(x)] = \sum_{i=1}^{3} \Pr_x[S_i] \, \mathbf{E}_x[q_u(x) - h(x)|S_i].$$

**Case 1:** In this case, the pointwise error is moderate – at most $(2 + \epsilon)$ – and we use gaussian anti-concentration to argue that the event has small probability mass. The event $H(x) \in [-\epsilon/Z, 0]$ implies that

$$\frac{w \cdot x - \theta}{Z} \in [-2a, 0] \implies q_u(x) \leq 1 + \epsilon \implies q_u(x) - h(x) \leq 2 + \epsilon,$$

using Item (3) in Theorem 3.5.

Since $h$ is $\epsilon$-regular, from Corollary 2.4 it follows that $\Pr_x[H(x) \in [-\epsilon/Z, 0]] \leq 3\epsilon$. So,

$$\Pr_x[S_1] \, \mathbf{E}_x[q_u(x) - h(x)|S_1] \leq (2 + \epsilon) \cdot 3\epsilon < 8\epsilon.$$

**Case 2:** This event has high probability, but in this range we get good pointwise approximation. The event $S_2$ implies that

$$H(x) \in [-1/2, 1/2] \setminus [-2a, 0] \implies q_u(x) \leq h(x) + \epsilon \implies q_u(x) - h(x) \leq \epsilon,$$

where we used Item (2)in Theorem 3.5. So,

$$\Pr_x[S_2] \, \mathbf{E}_x[q_u(x) - h(x)|S_2] \leq 1 \cdot \epsilon \leq \epsilon.$$

**Case 3:** Here we trade off the large magnitude of error (Item (4) in Theorem 3.5) with the small probability of the event (bounded by the Hoeffding bound). Define the intervals

$$I_j^+ = \left[\frac{j}{2}, \frac{(j+1)}{2}\right) \text{ for } j = 1, 2, \ldots$$

$$I_k^- = \left(\frac{-(k+1)}{2}, \frac{-k}{2}\right] \text{ for } k = 1, 2, \ldots.$$

We can write

$$\Pr_x[S_3] \mathbf{E}_x[q_u(x) - h(x)|S_3] = \sum_{j \geq 1} \Pr_x[H(x) \in I_j^+] \mathbf{E}_x[q_u(x) - h(x)|H(x) \in I_j^+]$$

$$+ \sum_{k \geq 1} \Pr_x[H(x) \in I_k^-] \mathbf{E}_x[q_u(x) - h(x)|H(x) \in I_k^-]. \quad (2)$$

Fix any integer $j \geq 1$. If $H(x) \in I_j^+$, then

$$\frac{j}{2} \leq H(x) < \frac{j+1}{2}.$$

Recalling that we have $|P(t)| \leq 2 \cdot (4t)^K$ for $t \geq 1/2$, we get that

$$q_u(x) = P(H(x)) \leq 2(2j+2)^K.$$

Since $h(x) = 1$, we get

$$q_u(x) - h(x) = q(x) - 1 \leq 2(2j+2)^K - 1. \quad (3)$$

Next we bound $\Pr_x[H(x) \in I_j^+]$ using the Hoeffding bound.

$$\Pr[H(x) \in I_j^+] \leq \Pr_x\left[w \cdot x - \theta \geq \frac{jZ}{2}\right] \leq \Pr_x\left[w \cdot x \geq \frac{jZ}{4}\right] \leq e^{-j^2 Z^2/32} \quad (4)$$

where the second inequality uses the fact that $|\theta| \leq Z/4$.

The analysis of the intervals $I_k^-$ is similar (except $h(x) = -1$). For $H(x) \in I_k^-$ we get

$$|H(x)| \leq \frac{k+1}{2} \implies q_u(x) \leq 2(k+1)^K \implies q_u(x) - h(x) \leq 2(2k+2)^K + 1. \quad (5)$$

Similarly, the Hoeffding bound gives

$$\Pr[H(x) \in I_k^-] \leq \Pr_x\left[w \cdot x - \theta \leq \frac{-kZ}{2}\right] \leq \Pr_x\left[w \cdot x \leq \frac{-kZ}{4}\right] \leq e^{-k^2 Z^2/32}. \quad (6)$$

Plugging equations (3), (4), (5), (6) back into (2), we get

$$\Pr_x[S_3] \, \mathbf{E}_x[q_u(x) - h(x)|S_3] \leq \sum_{j \geq 1} \frac{2(2j+2)^K - 1}{e^{j^2 Z^2/32}} + \sum_{k \geq 1} \frac{2(2k+2)^K + 1}{e^{k^2 Z^2/32}}$$

$$= 4 \sum_{j \geq 1} \frac{(2j+2)^K}{e^{j^2 Z^2/32}}$$

$$< 4 \sum_{j \geq 1} e^{j(2K - Z^2/32)}$$

where the last inequality follows by noting that, for $j \geq 1$, $(2j+2)^K < e^{2Kj}$ and $e^{j^2 Z^2/32} \geq e^{jZ^2/32}$. But now observe that

$$2K - \frac{Z^2}{32} < \frac{C \log^2(1/\epsilon)}{\epsilon^2} \left( 10c - \frac{C}{128} \right).$$

For a suitable choice of $C \gg c$, we have that $10c - C/128 \leq -1$, so

$$\Pr_x[S_3] \, \mathbf{E}_x[q_u(x) - h(x)|S_3] < 4 \sum_j e^{-jC \frac{\log^2(1/\epsilon)}{\epsilon^2}} < \epsilon.$$

Thus overall, we have $\mathbf{E}_x[q_u(x) - h(x)] \leq 10\epsilon$. $\qquad \square$

The lower sandwich bound follows by symmetry:

**Lemma 3.7.** $\mathbf{E}_x[h(x) - q_l(x)] < 10\epsilon$.

*Proof.* Since $q_l(x) \leq h(x)$ for every $x$, we also have $-h(x) \leq -q_l(x)$. Thus

$$-q_l(x) = P\left( \frac{\theta - w \cdot x}{Z} \right)$$

is an upper sandwich for the function $-h(x) = \text{sign}(\theta - w \cdot x)$. As this does not change the magnitude of $\theta$, we can apply the analysis of Lemma 3.6 to conclude that

$$\mathbf{E}_x[h(x) - q_l(x)] = \mathbf{E}_x[-q_l(x) - (-h(x))] < 10\epsilon.$$

$\qquad \square$

### 3.1.2 $|\theta|$ is large ($|\theta| > Z/4$)

We assume for simplicity that $\theta \geq Z/4$ (the case when $\theta$ is negative is handled similarly). The sandwich polynomials we use are:

$$r_u(x) = P\left( \frac{w \cdot x - Z/4}{Z} \right), \quad r_l(x) = -1. \tag{7}$$

**Lemma 3.8.** $h(x) \geq r_l(x)$ *for all* $x \in \{-1, +1\}^n$. *Further,* $\mathbf{E}_x[h(x) - r_l(x)] \leq 2\epsilon$.

*Proof.* Note that $\mathbf{E}_x[h(x) - r_l(x)] = 2 \Pr_x[h(x) = 1]$. For large enough $C$ we have $\Pr_x[h(x) = 1] = \Pr_x[w \cdot x \geq \theta] < e^{-Z^2/32} < \epsilon$. ∎

**Lemma 3.9.** $r_u(x) \geq h(x)$ *for all* $x \in \{-1, +1\}^n$. *Further,* $\mathbf{E}_x[r_u(x) - h(x)] \leq 12\epsilon$.

*Proof.* Observe that $r_u(x)$ is the upper sandwich polynomial for the halfspace $h'(x) = \text{sign}(w \cdot x - Z/4)$ as specified in Section 3.1.1. Thus we have

$$r_u(x) \geq h'(x) \geq h(x)$$

hence

$$\mathbf{E}_x[r_u(x) - h(x)] = \mathbf{E}_x[r_u(x) - h'(x)] + \mathbf{E}_x[h'(x) - h(x)].$$

By Lemma 3.6, $\mathbf{E}_x[r_u(x) - h'(x)] \leq 10\epsilon$ whereas by the Hoeffding bound $\mathbf{E}_x[h'(x) - h(x)] \leq 2\epsilon$ which completes the proof. ∎

# 4 Proof of Theorem 3.5

This section contains our proof of Theorem 3.5. The key step is to exhibit a low-degree univariate polynomial that approximates $\text{sign}(t)$ well when $|t| \in [a, 1]$ and is well-behaved even for larger values of $|t|$ to be compatible with the sandwich condition. We phrase this as a problem in univariate approximation. The solution we use is a low-degree polynomial $p(t)$ which is an optimal pointwise approximator to $\text{sign}(t)$ on $[-1, -a] \cup [a, 1]$. Such an optimal polynomial exists and we prove that it is well-behaved for large $|t|$, using ideas from classical approximation theory. However, it seems difficult to construct this polynomial explicitly and bound its error.

Recent work by [EY07] analyzes the error achieved by such a polynomial and in particular establishes the limiting behavior of the error function. For our purposes, though, we require the error to converge to the limit fairly rapidly and it is unclear whether the results of [EY07] guarantee this.

Instead, we bound the error by constructing a small error approximator $q(t)$ using Jackson's theorem together with standard amplification ideas. While $q(t)$ might not be well-behaved for large value of $t$, we only use it to bound from above the error of $p(t)$ on $[-1, -a] \cup [a, 1]$. Our approach has the advantage of being self-contained and elementary (using only standard ingredients from basic approximation theory) and matches the limiting bounds of [EY07] up to a constant factor.

For a bounded continuous function $f : [-1, 1] \to \mathbf{R}$, we define its *modulus of continuity* $\omega_f(\delta)$ as

$$\omega_f(\delta) := \sup\{|f(x) - f(y)| : x, y \in [-1, 1]; |x - y| \leq \delta\}.$$

A classical result of Dunham Jackson from the early twentieth century bounds the error of the best degree-$\ell$ approximation to $f$.

**Theorem 4.1. (Jackson's Theorem)** *[Che66] For $f$ as above and any integer $\ell \geq 1$, there exists a polynomial $J(t)$ with $\deg(J) \leq \ell$ so that*

$$\max_{t \in [-1,1]} |J(t) - f(t)| \leq 6\omega_f\left(\frac{1}{\ell}\right).$$

Recall the parameter $a = \frac{\epsilon^2}{C \log(1/\epsilon)}$ from the previous section. We now define the following parameter:

$$m := \frac{c \log(1/\epsilon)}{a}.$$

It will be crucial for us that $m$ is even (see in particular the last paragraph of the proof of Theorem 4.5.); for this condition to be satisfied, it is of course enough that $c$ is even. (We also note that the parameters $K$ and $m$ are such that $K = 4m + 2$.)

**Lemma 4.2.** *For $a, m$ as above, there is a polynomial $q(t)$ of degree at most $2m$ such that*

$$\max_{|t| \in [a,1]} |q(t) - \mathrm{sign}(t)| \leq \epsilon^2.$$

*Proof.* Define the continuous and piecewise linear function $f(x)$ as

$$f(x) = \begin{cases} \mathrm{sign}(t) & a \leq |t| \leq 1 \\ t/a & |t| \leq a. \end{cases}$$

Thus $f(x)$ increases linearly from $-1$ to $1$ in the range $[-a, a]$. A simple calculation shows that $\omega_f(\frac{1}{\ell}) = 1/(a\ell)$. Taking $\ell \geq 25/a$, Jackson's theorem implies the existence of a polynomial $J(t)$ of degree at most $\ell$ such that

$$\max_{a \leq |t| \leq 1} |J(t) - \mathrm{sign}(t)| \leq \max_{t \in [-1,1]} |J(t) - f(t)| \leq \frac{6}{a\ell} < \frac{1}{4}.$$

Our goal is to bring the error down to $\epsilon^2$. Rather than using Jackson's theorem for this (which would require degree $\tilde{O}(\epsilon^{-4})$), we use the degree-$k$ amplifying polynomial

$$A_k(u) := \sum_{j \geq \frac{k}{2}} \binom{k}{j} \left(\frac{1+u}{2}\right)^j \left(\frac{1-u}{2}\right)^{k-j}. \tag{8}$$

This polynomial has the following properties (easily proved via elementary calculation and also following from the Chernoff bound):

**Claim 4.3.** *The polynomial $A_k(u)$ satisfies:*

1. *If $u \in [3/5, 1]$, then $2A_k(u) - 1 \in [1 - 2e^{-k/6}, 1]$.*

2. *If $u \in [-1, -3/5]$, then $2A_k(u) - 1 \in [-1, -1 + 2e^{-k/6}]$.*

We define the polynomial

$$q(t) := 2A_k\left(\frac{4}{5}J(t)\right) - 1$$

where $k = 15\log(1/\epsilon)$. Scaling $J(t)$ by $\frac{4}{5}$ ensures that the argument to $A_k$ lies in the range $[-1, -3/5] \cup [3/5, 1]$ whenever $|t| \le a$. Applying Claim 4.3 with $k = 15\log(1/\epsilon)$ gives

$$\max_{|t| \in [a,1]} |q(t) - \text{sign}(t)| < 2e^{-k/6} < \epsilon^2.$$

Finally, by selecting $c$ large enough, we have

$$\deg(q) \le \deg(J)\deg(A_k) \le \frac{25}{a} \cdot 15\log(1/\epsilon) < \frac{2c}{a}\log(1/\epsilon) = 2m.$$

$\square$

We now present the "well-behaved" polynomial $p(t)$ mentioned at the beginning of this section. We will use Chebyshev's classical theorem on (weighted) real polynomial approximation ([Ach56], Chapter II).

**Theorem 4.4. (Chebyshev's Theorem)** *[Ach56] Let $f : [a,b] \to \mathbf{R}$ be a continuous function. Let $s : [a,b] \to \mathbf{R}$ be a continuous function that does not vanish on $[a,b]$. The polynomial $r(z)$ of degree $m$ that minimizes*

$$M(m) = \max_{t \in [a,b]} |f(z) - s(z)r(z)|$$

*is unique, and it is characterized by the property that there exist $m+2$ points $a \le z_0 < z_1 \cdots < z_{m+1} \le b$ such that for each $z_i$*

$$M(m) = |f(z_i) - s(z_i)r(z_i)|$$

*and the sign of the error at the $z_i$'s alternates.*

**Theorem 4.5.** *Let $a$ and $m$ be as specified in Section 3. There is a univariate polynomial $p(t)$ where $\deg(p) \le 2m+1$ such that:*

1. *$p(t) \in [\text{sign}(t) - \epsilon^2, \text{sign}(t) + \epsilon^2]$ for all $|t| \in [a,1]$;*

2. *$p(t) \in [-(1+\epsilon^2), 1+\epsilon^2]$ for all $t \in [-a,a]$;*

3. *$p(t)$ is monotonically increasing on the intervals $(-\infty, -1]$ and $[1, \infty)$.*

*Proof.* The polynomial $p(t)$ is a best uniform approximation (such an approximation is guaranteed to exist [Riv74]) to the function $\text{sign}(t)$ of degree at most $2m+1$ over the domain $[-1, -a] \cup [a, 1]$. Applying Lemma 4.2, we get

$$\max_{|t| \in [a,1]} |p(t) - \text{sign}(t)| \le \max_{|t| \in [a,1]} |q(t) - \text{sign}(t)| \le \epsilon^2$$

which gives Property (1).

We can assume that $p(t)$ is odd (by replacing it with $(p(t) - p(-t))/2$ if needed). So we can write $p(t) = t \cdot r(t^2)$, where $r(z)$ is a polynomial of degree $m$ that minimizes

$$M(m) = \min_{r:\ \deg(r) \le m} \sup_{z \in [a^2, 1]} |1 - \sqrt{z} r(z)|.$$

Invoking Chebyshev's theorem with $f(z) = 1$ and $s(z) = \sqrt{z}$ (which does not vanish on $[a^2, 1]$), we infer that the optimal polynomial $r(z)$ of degree $m$ is unique and it has an *alternating* sequence of points

$$a^2 \le z_0 < z_1 \ldots < z_{m+1} \le 1$$

so that the error $1 - \sqrt{z} r(z)$ achieves its maximum magnitude exactly at the points $z_i$, and the sign of the error alternates.

Set $t_i = \sqrt{z_i} > 0$ so that

$$a \le t_0 < t_1 \ldots < t_{m+1} \le 1.$$

Let $\phi(t)$ be the error function $\phi(t) = p(t) - \text{sign}(t)$. Note that for $t \ge a$, we have

$$\phi(t) = p(t) - 1,$$
$$\phi(-t) = p(-t) - (-1) = -p(t) + 1 = -\phi(t).$$

For each $t_i$, we have

$$|\phi(t_i)| = |\phi(-t_i)| = M(m).$$

Now consider the interval $[a, 1]$, on which $\phi(t) = p(t) - 1$. Note that $\phi'(t)$ is well defined and equals $p'(t)$ at any point in $(a, 1)$. The points $t_1, \ldots, t_m$ lie in $(a, 1)$ and they are local maxima/minima, since $\phi(t)$ cannot increase in magnitude in the neighborhood of $t_i$. Thus $\phi'(t_i) = p'(t_i) = 0$ for each $i \in [m]$. Similarly, we can show that $\phi'(-t_i) = p'(-t_i) = 0$ for $i \in [m]$. But $\deg(p')$ is at most $2m$, and so we have located all its roots. As we now show, this allows us to determine the sign of $p$ in the intervals $[-\infty, -1], [-a, a]$ and $[1, \infty]$.

Note that $p(t_1)$ is close to 1 whereas $p(-t_1)$ is close to $-1$, and thus $p$ increases monotonically in the interval $(-t_1, t_1)$ which includes $[-a, a]$. Also $t_1$ is a local maximum for $p$, which shows that the $t_i$s are maxima when $i$ is odd, and minima when $i$ is even. Thus, since $m$ is even, $p(t_m)$ is a local minimum, so $p(t)$ increase monotonically in the range $(t_m, \infty)$, which includes $[1, \infty)$. Whereas $-t_1$ is a local minimum for $p$, so $p(-t_i)$ are local minima for odd $i$ and maxima for even $i$, hence $p(t)$ is monotonically increasing in the range $(-\infty, t_m)$ which contains $(-\infty, -1]$. □

Using the polynomial $p(t)$, we now construct the polynomial $P(t)$ which is a good "upper" approximator to $\text{sign}(t)$ (i.e. $P(t) \ge \text{sign}(t)$ for all $t$), thus completing the proof of Theorem 3.5.

To help the reader visualize $p(t)$, we provide a schematic representation in Figure 2. (We remark that, as before, this figure is not an actual plot, but rather is intended to qualitatively illustrate the guarantees on the behavior of $p$ on various intervals.)

Let us recall the statement of Theorem 3.5:

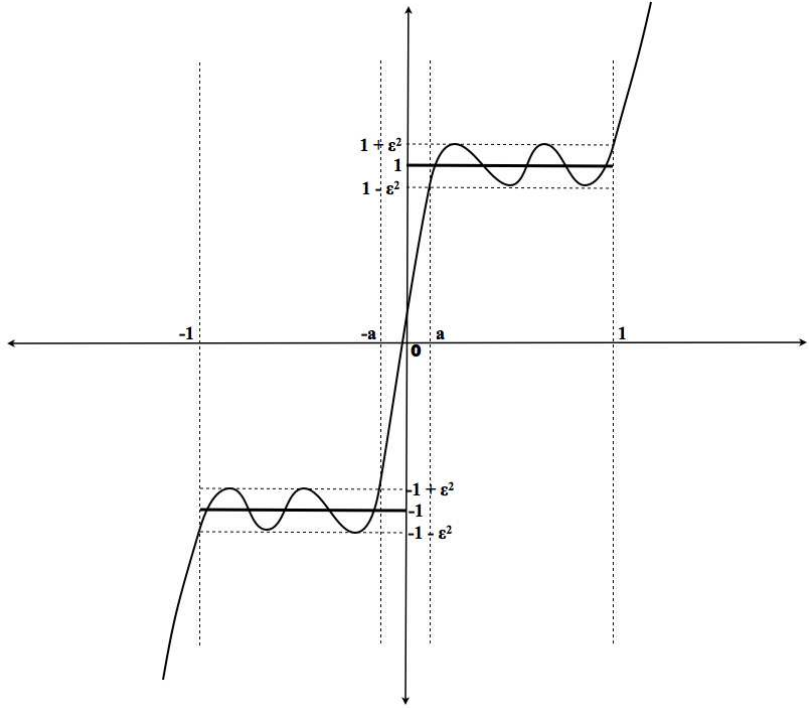

Figure 2: Qualitative representation of polynomial $p$.

**Theorem 3.5.** *(Restated.)* Let $0 < \epsilon < 0.1$ and let $a$ and $K$ be as defined above. There is a univariate polynomial $P(t)$ such that $\deg(P) \leq K$ with the following properties:

(1) $P(t) \geq \text{sign}(t) \geq -P(-t)$ for all $t \in \mathbf{R}$;

(2) $P(t) \in [\text{sign}(t), \text{sign}(t) + \epsilon]$ for $t \in [-1/2, -2a] \bigcup [0, 1/2]$;

(3) $P(t) \in [-1, 1 + \epsilon]$ for $t \in (-2a, 0)$;

(4) $|P(t)| \leq 2 \cdot (4t)^K$ for all $|t| \geq 1/2$.

*Proof.* Let $p$ denote the polynomial of degree $2m + 1$ from Theorem 4.5. Consider the following polynomial:

$$P(t) = \frac{1}{2}(1 + \epsilon^2 + p(t + a))^2 - 1.$$

Note that $\deg(P) = 2 \deg(p) \leq K$. We now consider the behavior of $P$ on the relevant intervals. We repeatedly use the inequality

$$\frac{1}{2}(2 + 2\epsilon^2)^2 - 1 = 1 + 4\epsilon^2 + 2\epsilon^4 \leq 1 + \epsilon$$

which holds since $\epsilon < \frac{1}{10}$. Note that $P(t) \geq -1$ holds for all $t$. We now analyze the behavior of $P(t)$ interval by interval:

(a) $t \in [-1-a, -2a]$. Here $p(t+a) \in [-1-\epsilon^2, -1+\epsilon^2]$, hence $P(t) \in [-1, -1+\epsilon]$.

(b) $t \in (-2a, 0)$. Here $p(t+a) \in [-1-\epsilon^2, 1+\epsilon^2]$, hence $P(t) \in [-1, 1+\epsilon]$.

(c) $t \in [0, 1-a]$. Here $p(t+a) \in [1-\epsilon^2, 1+\epsilon^2]$, hence $P(t) \in [1, 1+\epsilon]$.

(d) $t \in (1-a, \infty]$. Here $p(t+a) \geq 1-\epsilon^2$, hence $P(t) \geq 1$.

This shows that $P(t) \geq \text{sign}(t)$ for all $t \in \mathbf{R}$. Thus we also have

$$P(-t) \geq \text{sign}(-t) \ \Rightarrow \ \text{sign}(t) \geq -P(-t)$$

which establishes Property (1). Properties (2) and (3) follow immediately from (a), (b) and (c) above.

For Property (d), we use the following standard fact from approximation theory.

**Fact 4.7.** [Riv74] Let $a(t)$ be a polynomial of degree at most $d$ for which $|a(t)| \leq b$ in the interval $[-1, 1]$. Then $|a(t)| \leq b|2t|^d$ for all $|t| \geq 1$.

Taking $a(t)$ to be $P(t/2)$, properties (2) and (3) give us that $|P(t/2)| \leq 2$ for $t \in [-1, 1]$. So the fact gives $|P(t/2)| < 2|2t|^{4m+2}$ for $|t| \geq 1$, i.e. $|P(t)| < 2|4t|^{4m+2}$ for $|t| \geq 1/2$. □

# 5 Fooling non-regular halfspaces

In this section we show how to fool halfspaces that are not regular. We proceed by case analysis based on the *critical index* of the halfspace, which we define shortly. Throughout this section we assume that the weights of the halfspace are decreasing:

$$|w_1| \geq |w_2| \ldots \geq |w_n|.$$

We can assume this without loss of generality because we are going to prove that, for a suitable $k$, *any* $k$-wise independent distribution fools such halfspaces, and the property of being $k$-wise independent is clearly invariant under permutation of the variables.

Some notation: For $T \subseteq [n]$ we denote by $\sigma_T$ the quantity $\sigma_T := \sqrt{\sum_{i \in T} w_i^2}$. For $k \in [n]$ we also write $\sigma_k$ for $\sigma_{\{k,k+1,\ldots,n\}}$.

**Definition 5.1** (Critical index). We define the *$\tau$-critical index $\ell(\tau)$ of a halfspace $h =$* $\text{sign}(w \cdot x - \theta)$ as the smallest index $i \in [n]$ for which

$$|w_i| \leq \tau \cdot \sigma_i.$$

If this inequality does not hold for any $i \in [n]$, we define $\ell(\tau) = \infty$.

Note that a halfspace is $\tau$-regular if $\ell(\tau) = 1$; in this section we handle the case $\ell(\tau) > 1$.

We assume without loss of generality that $\epsilon$ is sufficiently small. Given $\epsilon$, our threshold for the critical index is

$$L(\epsilon) := \frac{8 \log^2(10/\epsilon)}{\epsilon^2}.$$

We argue separately depending on whether $\ell(\epsilon) > L(\epsilon)$ or not. Both proofs rely on the following simple property of $k$-wise independent distributions.

**Fact 5.2.** Let $\mathcal{D}$ be a $k$-wise independent distribution over $\{-1, +1\}^n$. Condition on any fixed values for any $t \leq k$ bits of $\mathcal{D}$, and let $\mathcal{D}'$ be the projection of $\mathcal{D}$ on the other $n - t$ bits. Then $\mathcal{D}'$ is $(k - t)$-wise independent.

The first theorem addresses the simpler case when $\ell(\epsilon) \leq L(\epsilon)$.

**Theorem 5.3** (Fooling non-regular halfspaces with small critical index). *Let $h(x)$ be a halfspace with $\epsilon$-critical index $\ell(\epsilon) \leq L(\epsilon)$. Then any $(K(\epsilon) + L(\epsilon))$-wise independent distribution $O(\epsilon)$-fools $h$.*

*Proof.* Condition on any setting to the first $\ell - 1$ variables. Each of these defines a halfspace of the form

$$h'(x) = \text{sign} \left( \sum_{i \geq \ell} w_i x_i - \theta' \right)$$

where $\theta'$ depends on the values assigned to the head. Every such halfspace is $\epsilon$-regular by the definition of $\epsilon$-critical index. Also, the conditional distribution on the remaining variables is $K(\epsilon)$-wise independent by Fact 5.2. Thus, Theorem 3.2 implies that we fool $h'$ with error $\epsilon$. Since both the uniform distribution and $\mathcal{D}$ induce the same (uniform) distribution on the first $\ell - 1$ variables, an averaging argument concludes the proof of the theorem. □

In the rest of this section we study the case of large critical index $\ell(\epsilon) > L(\epsilon)$, and prove the following theorem.

**Theorem 5.4** (Fooling non-regular halfspaces with large critical index). *Let $h(x)$ be a halfspace with critical index $\ell(\epsilon) > L(\epsilon)$. Any $(L(\epsilon) + 2)$-wise independent distribution $\mathcal{D}$ fools $h$ with error $9\epsilon$.*

To prove Theorem 5.4 we partition the coordinate set $[n]$ into a *head $H$* consisting of the first $L(\epsilon)$ coordinates, and a *tail $T = [n] \setminus H$* consisting of the rest. We then show that a random setting of the head variables induces with high probability a partial sum $\sum_{i \in H} w_i x_i - \theta$ which is so large in magnitude that the values of the tail variables are essentially irrelevant, in the sense that they are very unlikely to change the sign of $w \cdot x - \theta$ and hence the value of the halfspace.

We will show that this statement holds both for the uniform distribution and for the distribution $\mathcal{D}$ with limited independence. For the latter we will use that after restricting the variables in the head we still have a 2-wise independent distribution on the tail (by Fact 5.2), which is enough for Chebyshev's concentration bound to apply. To show that the partial sum is likely to be large we use ideas from [Ser07], in particular that the weights decrease geometrically up to the critical index.

## 5.1 Proof of Theorem 5.4

We partition the coordinate set $[n]$ into a *head $H$* consisting of the first $L(\epsilon)$ coordinates, and a *tail $T = [n] \setminus H$* consisting of the rest. Any fixing of the variables in $H$ results in a halfspace

$$h'(x_T) := \text{sign}\left(\sum_{i \in T} w_i x_i - \theta'_H\right)$$

over the tail variables $x_T$ where

$$\theta'_H := \theta - \sum_{i \in H} w_i x_i.$$

As discussed before, our goal is to show that, for a random setting of the head variables, $\theta'_H$ is likely to be so large in magnitude that the value of the tail sum $\sum_{i \in T} w_i x_i$ is unlikely to influence the outcome of $h(x)$. The key idea here is the following lemma from [Ser07] showing that the weights decrease geometrically up to the critical index.

**Lemma 5.5.** *For any $1 \le i < j \le \ell + 1$ we have*

$$|w_j| \le \sigma_j < \left(\sqrt{1 - \epsilon^2}\right)^{j-i} \sigma_i \le \left(\sqrt{1 - \epsilon^2}\right)^{j-i} |w_i|/\epsilon.$$

*In particular, if $j \ge i + (4/\epsilon^2)\ln(1/\epsilon)$ then*

$$|w_j| \le |w_i|/3.$$

*Proof.* For any $k \le \ell$, we have by the definition of $\epsilon$-critical index that

$$w_k^2 > \epsilon^2 \sigma_k^2.$$

Hence

$$\sigma_{k+1}^2 = \sigma_k^2 - w_k^2 < (1 - \epsilon^2)\sigma_k^2.$$

Repeating this calculation yields

$$\sigma_j^2 < (1 - \epsilon^2)^{j-i} \sigma_i^2.$$

To conclude the first chain of inequalities in the statement of the lemma, use again $\sigma_i^2 < w_i^2/\epsilon^2$ and the obvious inequality $\sigma_j^2 \ge w_j^2$. The "in particular" part can be verified by straightforward calculation, using that $\epsilon$ is sufficiently small. □

Now consider the set of
$$t := \log(10/\epsilon)$$
"nicely separated" coordinates (variables)
$$G := \{k_i := 1 + i \cdot (4/\epsilon^2) \ln(1/\epsilon) : i = 0, 1, \ldots, t - 1\} \subseteq H.$$

Observe that indeed $G \subseteq H$ because the maximum index in $G$ is at most $1 + t \cdot (4/\epsilon^2) \log(1/\epsilon) \leq (4/\epsilon^2) \log^2(10/\epsilon)$, whereas $H$ consists of all the first $L(\epsilon) = (8/\epsilon^2) \log^2(10/\epsilon)$ indices. The key features of $G$ are that we can apply the 'in particular' part of Lemma 5.5 and prove the following claim.

**Claim 5.6.** $\sigma_T < \epsilon |w_{k_t}|$.

*Proof.* By our choice of $L(\epsilon), t$, and $k_t$, we have
$$L(\epsilon) - k_t \geq 8 \log^2(10/\epsilon)/\epsilon^2 - 4 \log^2(10/\epsilon)/\epsilon^2 \geq \log^2(1/\epsilon)/\epsilon^2.$$

An application of Lemma 5.5 gives
$$\sigma_T < \sqrt{1 - \epsilon^2}^{\log^2(1/\epsilon)/\epsilon^2} |w_{k_t}|/\epsilon \leq \epsilon^2 |w_{k_t}|/\epsilon = \epsilon |w_{k_t}|$$
where we use that $\epsilon$ is sufficiently small. $\qquad\square$

We now show that a random setting of $H$ is likely to result in a value of $|\theta'_H|$ which is at least $|w_{k_t}|/4$. The proof relies on the following claim.

**Claim 5.7.** *Let $v_1 > v_2 > \cdots > v_t > 0$ be a sequence of numbers so that $v_{i+1} \leq v_i/3$. Then for any two points $x \neq y \in \{-1, +1\}^t$, we have $|v \cdot x - v \cdot y| \geq v_t$.*

*Proof.* Let $z := x - y \in \{-2, 0, 2\}^t$, which is not zero. Let $j \leq t$ be the smallest index such that $z_j \neq 0$. Then
$$|v \cdot x - v \cdot y| = |v \cdot z| = |\sum_{i \geq j} v_i z_i| \geq |v_j z_j| - \sum_{i > j} |v_i z_i| \geq 2(v_j - \sum_{i > j} v_i)$$
$$\geq 2(v_j - \sum_{i > j} \frac{v_j}{3^{i-j}}) \geq 2(v_j - v_j/2) = v_j \geq v_t,$$
using $v_i \leq v_j/3^{i-j}$ by assumption. $\qquad\square$

We are now ready to show our intended lemma:

**Lemma 5.8.** $\Pr_{x_i : i \in H} \left[ |\theta - \sum_{i \in H} w_i x_i| \leq |w_{k_t}|/4 \right] \leq \epsilon/10.$

*Proof.* Fix any assignment to the variables in $H \setminus G$. For this fixing, the event $|\theta - \sum_{i \in H} w_i x_i| \leq |w_{k_t}|/4$ happens only if

$$\sum_{i \in G} w_i x_i \in \left[ \theta - \sum_{i \in H \setminus G} w_i x_i - |w_{k_t}|/4, \theta - \sum_{i \in H \setminus G} w_i x_i + |w_{k_t}|/4 \right],$$

i.e., $\sum_{i \in G} w_i x_i$ falls in an interval of length $|w_{k_t}|/2$. Applying Claim 5.7 to the weights in $G$, any two possible outcomes of $\sum_{i \in G} w_i x_i$ differ by at least $|w_{k_t}|$. So there is at most one setting $x_{k_1} = a_1, \ldots, x_{k_t} = a_t$ of the variables in $G$ for which this event occurs. This setting has probability at most $2^{-t} = \epsilon/10$. □

With this lemma in hand, we can show that limited independence suffices to fool halfspaces with a large critical index.

*Proof of Theorem 5.4.* We compare the behavior of $h(x)$ on $\mathcal{D}$ and the uniform distribution $\mathcal{U}$. In either case, the marginal distribution for the variables in $H$ is uniform. For each setting of these variables, we are left with a halfspace of the form $h'(x_T) = \text{sign}(\sum_{i \in T} w_i x_i - \theta'_H)$ on the variables in $T$. The combination of Lemma 5.8 and Claim 5.6 shows that with probability at least $1 - \epsilon/10$ we have

$$\left| \theta - \sum_{i \in H} w_i \cdot x_i \right| \geq \frac{|w_{k_t}|}{4} \geq \frac{\sigma_T}{4\epsilon}. \qquad (\star)$$

We condition on this event $(\star)$. Consider the projections $\mathcal{U}'$ and $\mathcal{D}'$ of $\mathcal{U}$ and $\mathcal{D}$ on $x_T$. By Fact 5.2, $\mathcal{D}'$ is 2-wise independent. We now argue that for both $\mathcal{U}'$ and $\mathcal{D}'$, it is very likely that $h'(x_T) = -\text{sign}(\theta'_H)$ (for small enough $\epsilon$). Indeed if this does not happen, then we have

$$\left| \sum_{i \in T} w_i x_i \right| \geq \left| \theta - \sum_{i \in H} w_i \cdot x_i \right| \geq \frac{|w_{k_t}|}{4} \geq \frac{\sigma_T}{4\epsilon}.$$

Under the uniform distribution, by a Hoeffding bound (Theorem 2.5), the probability of this event is bounded by

$$\Pr_{x \sim \mathcal{U}'} \left[ \left| \sum_{i \in T} w_i x_i \right| \geq \frac{\sigma_T}{4\epsilon} \right] \leq 2e^{-\frac{1}{32\epsilon^2}} \ll 4\epsilon.$$

While by Chebyshev's inequality (Theorem 2.6) we get

$$\Pr_{x \sim \mathcal{D}'} \left[ \left| \sum_{i \in T} w_i x_i \right| \geq \frac{\sigma_T}{4\epsilon} \right] \leq 16\epsilon^2 \leq 4\epsilon.$$

Thus, we have

$$| \mathbf{E}_{\mathcal{D}'}[h'(x_T)] - \mathbf{E}_{\mathcal{U}'}[h'(x_T)]| \leq 2| \Pr_{\mathcal{D}'}[h'(x_T) = -\text{sign}(\theta'_H)] - \Pr_{\mathcal{U}'}[h'(x_T) = -\text{sign}(\theta'_H)]| \leq 8\epsilon.$$

To conclude, our goal was to bound from above $|\mathbf{E}_{\mathcal{U}}[h(x)] - \mathbf{E}_{\mathcal{D}}[h(x)]|$. Using the fact that both distributions induce the uniform distribution on variables in $H$, and conditioning on the event $(\star)$, we get

$$|\mathbf{E}_{\mathcal{U}}[h(x)] - \mathbf{E}_{\mathcal{D}}[h(x)]| \leq 8\epsilon + 2 \cdot \epsilon/10 < 9\epsilon.$$

$\square$

Our main result, Theorem 1.2, follows immediately from Theorem 3.2, Theorem 5.3 and Theorem 5.4.

## 5.2 Proof of the main theorem

For completeness in this section we summarize what is needed to prove our main theorem.

**Theorem 1.2** (Main). *(Restated.)* Let $\mathcal{D}$ be a $k$-wise independent distribution on $\{-1, +1\}^n$, and let $h : \{-1, +1\}^n \to \{-1, +1\}$ be a halfspace. Then $\mathcal{D}$ fools $h$ with error $\epsilon$, i.e.,

$$|\mathbf{E}_{x \leftarrow \mathcal{D}}[h(x)] - \mathbf{E}_{x \leftarrow \mathcal{U}}[h(x)]| \leq \epsilon, \qquad \text{provided} \qquad k \geq \frac{C}{\epsilon^2} \log^2\left(\frac{1}{\epsilon}\right),$$

where $C$ is an absolute constant and $\mathcal{U}$ is the uniform distribution over $\{-1, +1\}^n$.

*Proof.* Consider the parameters $K(\epsilon), L(\epsilon)$ defined in Sections 3 and 5, respectively, and recall that they are both $O(\log^2(1/\epsilon)/\epsilon^2)$. For a given halfspace, consider its critical index $\ell$. If $\ell \leq L(\epsilon)$ we apply Theorem 5.3, otherwise we apply Theorem 5.4. $\square$

# 6 Conclusions

We feel that Theorem 1.2 is of independent interest and may find other applications aside from pseudorandomness. For instance, consider the problem of estimating the influence of a variable in a halfspace [GR09]. It is easy to verify that for any halfspace $h(x) = \text{sign}(\sum_{i=1}^n w_i x_i - \theta)$ the influence of the $i$-th variable equals $\mathbf{E}_{y \leftarrow \mathcal{U}}[h'(y)]$, where $h'$ is the halfspace defined by $h'(y) = \text{sign}(\sum_{j,j \neq i} w_j y_j - \theta y_i + w_i)$. Thus, one can use $\tilde{O}(\epsilon^{-2})$-wise independence to estimate the influence to within an additive $\epsilon$. Note that, for any halfspace, the bias and the influences together are (respectively) the Fourier coefficients at levels 0 and 1. They are collectively called the Chow parameters of a halfspace (after a theorem of C.K. Chow showing that these numbers uniquely specify the halfspace [Cho61]) and have been well-studied in the literature [Gol06, Ser07, OS08, MORS09]. Our result implies that the Chow parameters of a halfspace can be estimated to within accuracy $\epsilon$ using bounded independence.

Our results, together with the lower bound of [BGGP07], are essentially optimal in terms of characterizing the degree of independence that is required to $\epsilon$-fool halfspaces. However, many natural and interesting directions remain for future work.

One obvious goal is to construct unconditional pseudorandom generators for halfspaces that have a better dependence on $\epsilon$ than our construction. The ultimate goal here is to achieve the information-theoretic optimal possible seed length, i.e. $s = O(\log(n/\epsilon))$.

Another natural, though perhaps challenging, goal is to understand the degree of independence that is required to $\epsilon$-fool degree-$d$ polynomial threshold functions over $\{-1, +1\}^n$. For constant $\epsilon$ and constant $d$, does $\Theta(1)$-wise independence suffice to fool degree-$d$ PTFs? As far as we know nothing is known about this question, even for $d = 2$.

A third question that is related to our work is whether it is possible to derandomize the problem of approximately counting the number of satisfying assignments for a given halfspace. Our results give a single fixed and easily constructible set of $n^{\tilde{O}(1/\epsilon^2)}$ many points which can be used to deterministically obtain a $\pm\epsilon$-accurate estimate of $\Pr_{\mathcal{U}}[f(x) = 1]$ for any halfspace $f$ in time $n^{\tilde{O}(1/\epsilon^2)}$. However, there is a deterministic algorithm of [Ser07] which takes integer weights and threshold $w_1, \ldots, w_n, \theta$ (each poly($n$) bits long) as input and runs in time poly($n$) $\cdot 2^{\tilde{O}(1/\epsilon^2)}$. Can a poly($n, 1/\epsilon$)-time deterministic algorithm be obtained?

**Acknowledgements.** Rocco Servedio thanks Troy Lee for a helpful conversation about amplifying polynomials and Adam Klivans for useful conversations about Jackson's Theorem. Parikshit Gopalan would like to thank Jaikumar Radhakrishnan and Amir Shpilka for many stimulating discussions about this problem.